# Veselago lensing with ultracold atoms in an optical lattice

Martin Leder, Christopher Grossert, and Martin Weitz

*Institut für Angewandte Physik der Universität Bonn, Wegelerstraße 8, 53115 Bonn, Germany*

**Veselago pointed out that electromagnetic wave theory allows for materials with a negative index of refraction, in which most known optical phenomena would be reversed. A slab of such a material can focus light by negative refraction, an imaging technique strikingly different from conventional positive refractive index optics, where curved surfaces bend the rays to form an image of an object. Here we demonstrate Veselago lensing for matter waves, using ultracold atoms in an optical lattice. A relativistic, i.e. photon-like, dispersion relation for rubidium atoms is realized with a bichromatic optical lattice potential. We rely on a Raman π-pulse technique to transfer atoms between two different branches of the dispersion relation, resulting in a focusing completely analogous to the effect described by Veselago for light waves. Future prospects of the demonstrated effects include novel sub-de Broglie wave imaging applications.**

Veselago lensing is a concept based on negative refraction[1-3], where a spatially diverging pencil of rays is focused to a spatially converging one during the transition from a medium with positive refractive index to a negative index material, as depicted in Fig.1. Pendry has shown that the spatial resolution of a negative index lens is not limited by the usual diffraction limit, because it restores the evanescent waves[4]. Using a thin silver film, a spatial resolution below the diffraction limit has indeed been experimentally observed[5], and other experiments have demonstrated negative refractive index optics with optical metamaterials[6]. Veselago lensing has also been proposed for electrons in graphene material[7] and cold atoms in dark state media[8], but so far no experimental realization has been reported. The refractive action of a Veselago lens differs clearly from usual lenses, for the latter see also earlier work on the lensing of atomic de Broglie waves using light potentials with quadratic dispersion[9,10]. The essential prerequisite to transfer such novel light optics effects towards matter waves is a relativistic, i.e. photon-like, dispersion relation, an issue that has been achieved in some recent experiments with cold atoms in optical lattices, including the observation of Klein tunneling, the moving and creation of Dirac points, and Zitterbewegung[11-13], see also refs 7,14-16.

Here we demonstrate Veselago lensing for matter waves. Our experiment uses ultracold rubidium atoms in a bichromatic optical lattice with a linear, i.e. relativistic, dispersion relation. A Raman π-pulse transfers the atomic de Broglie waves between different branches of the dispersion relation, and the relativistic lensing occurs by a backwards propagation of atomic wavepackets on an energetically mirrored branch of the dispersion relation. We study negative refraction and Veselago lensing both in a one-dimensional geometry and perform a ray-tracing simulation of a two-dimensional Veselago lens.

**Results**

*Experimental Background*

To tailor the atomic dispersion relation to resemble that of light in positive and negative index materials respectively, we modify the free atomic dispersion by adding a suitable periodic potential induced by laser fields. They imprint a bichromatic optical lattice potential of the form $V(x) = (V_1/2)\cos(2\kappa x) + (V_2/2)\cos(4\kappa x + \varphi)$, where $V_1$ denotes the potential depth of a usual standing wave lattice potential of spatial periodicity $\lambda/2$ and $V_2$ that of a higher lattice harmonic of spatial periodicity $\lambda/4$, generated by the dispersion of multiphoton Raman transitions[17]. Here, $\lambda$ denotes the laser wavelength with wavevector $\kappa = 2\pi/\lambda$, and $\varphi$ the relative phase between lattice harmonics. For a relative phase $\varphi=180^0$ and a suitable choice of the amplitude of lattice harmonics, the splitting between the first and the second excited Bloch band vanishes and the corresponding dispersion relation near the crossing becomes linear, i.e. relativistic, see the solid line in Fig.2. This is understood from the destructive interference of the contributions from first order Bragg scattering of the higher periodicity lattice and second order Bragg scattering of the standing wave lattice. In general, if we set the zero of the energy scale to the position of the crossing, the energy of atomic eigenstates near the crossing is $E = \pm\sqrt{(c_{eff}\hbar k_x)^2 + (\Delta E/2)^2}$, where $\hbar k_x$ denotes the atomic quasimomentum along the lattice axis, and the splitting between band is $\Delta E = \left|(V_1^2/32E_r) + (V_2/2)e^{i\varphi}\right|$ in the limit $|k_x| \ll \kappa$ and $\Delta E \ll E_r$, with $E_r = (\hbar\kappa)^2/(2m)$ as the recoil energy. Further, $c_{eff} = 2\hbar\kappa/m$ is an effective light speed, which for rubidium atoms and a 783nm lattice laser wavelength is 1.1cm/s, ten orders of magnitude smaller than the speed of light in vacuum. The dynamics of atoms in the bichromatic lattice near the crossing is well described by a one-dimensional Dirac equation, which allows us to study relativistic wave equation predictions with ultracold atoms[18]. By tuning of the splitting ΔE, moreover different projections of a two-dimensional Dirac cone can be realized.

*One-dimensional Veselago lensing of matter waves*

We have studied the one-dimensional analog of Veselago lensing with cold atoms in the optical lattice, in the presence of an additional, spatially slowly varying potential. In a single dimension, two rays are possible (atomic paths along and opposed to the x-axis of Fig.1 respectively), and Fig.3a shows a schematic of the relevant part of the atomic dispersion relation for the case of a vanishing splitting between bands, with $E(k_x) = \pm \hbar c_{eff} |k_x|$. The atoms initially propagate in the lattice with linear dispersion relation with an energy-independent velocity of $\pm c_{eff}$ respectively, which above the crossing, i.e. for E>0, in analogy to light optics can well be described by propagation in a medium with positive refractive index, say n=1. Due to the linear dispersion relation, the atoms remain at a group velocity of $\pm c_{eff}$ even in the presence of the additional (here confining) potential. Transitions to states below the crossing, i.e. with E<0, are induced with a short (broadband) four-photon Raman π-pulse. This reverses the path of propagation, resulting in spatially converging atom paths. The group velocity is now opposed to the wavevector $k_x$, and we conclude that the regime with E<0 corresponds to propagation in a medium with negative refractive index, n=-1. The π-pulse reverses the temporal evolution, similar as in a spin-Echo experiment[19], and after the second propagation time the paths close in a focus.

In our experiment, a Bose-Einstein condensate of rubidium ($^{87}$Rb) atoms in the $m_F=0$ spin projection of the F=1 ground state hyperfine component is loaded into a weakly confining optical dipole trapping potential with a 70Hz radial trapping frequency (see Supplementary information). The atoms are irradiated with a short pulse performed with optical beams simultaneously coupling the two Raman transitions $|m_F=0,\hbar k_x\rangle \rightarrow |m_F=-1,\pm 2\hbar\kappa+\hbar k_x\rangle$ respectively, after which the lattice beams are activated. This leaves the atomic cloud in the lattice with a coherent superposition of two spatially diverging atomic paths of "quasi-relativistic" velocities, and energies tuned slightly above the position of the crossing shown in Fig.3a. Fig.3b (left) shows a series of absorption images for different propagation times in the lattice. After an expansion time of 1.3ms, a four-photon Raman π-pulse coupling the states $|m_F=-1,-2\hbar\kappa+\hbar k_x\rangle$ and $|m_F=-1,2\hbar\kappa+\hbar k_x\rangle$ is applied, which causes the spatially diverging atom paths to converge. The paths meet after a second propagation time, and form an image of the initial atomic cloud (Due to the used coherent preparation of the two rays, one here also expects an atomic interference pattern[20,21], of spatial periodicity λ/4, which is below our 3.8μm experimental imaging resolution.). Fig.3c shows the variation of the total width of the atomic distribution versus time, which again clearly shows the focusing of the spatially expanding cloud subject to the relativistic dispersion relation in the lattice by the π-pulse. The observed 16μm width of the image formed by the one-dimensional Veselago lens at ≅2.3ms propagation time equals the initial cloud size within experimental uncertainties. For comparison, Figs.3d and 3e show experimental data recorded for a non-vanishing energy gap ΔE between the two Bloch bands, where the image is clearly larger than the initial cloud size. This is attributed to the atomic velocity now depending on energy, as in[9,10], which does not correspond to the case of Veselago lensing. The energy spread acquired from the spatially varying dipole potential here causes different group velocities, which results in larger aberrations. The Veselago lens, see the data shown in Fig.3c, achieves sharp refocusing even in the presence of the additional potential, as is understood from the (in an ideal experimental situation) energy independence of the group velocity. Our experimental findings are supported by numerical simulations (see methods: numerical calculations).

*Ray tracing simulation of two-dimensional relativistic lensing*

We also carried out a ray-tracing simulation of a two-dimensional Veselago lens. The two-dimensional optical dispersion is $E(k_x, k_y) = \pm \hbar c_{eff} \sqrt{k_x^2 + k_z^2}$ for a n=±1 refractive index respectively, where the group velocity $\vec{v}_g = \pm c_{eff} \vec{k}/|\vec{k}|$ is parallel (antiparallel) to the wavevector $\vec{k}$ in the positive (negative) energy branch respectively. Our one-dimensional optical lattice is used to realize different cuts along the x-axis of the two-dimensional Dirac cone, with $\Delta E = c_{eff} 2\hbar k_z$. The optical propagation along the z-axis is here replaced by time. The corresponding transformation for a ray with angle α to the optical axis is $t(\alpha) = \dfrac{z}{c_{eff} \cos(\alpha)}$, which ensures that rays emitted from the source under different angles α reach the Veselago lens at a fixed value of z=L, see Fig.1. The refraction induced by a four-photon Raman π-pulse between momentum states along the lattice axis of $|-2\hbar\kappa+\hbar k_x\rangle$ and $|2\hbar\kappa+\hbar k_x\rangle$ conserves the wavevector component orthogonal to the lens surface $k_x$ in the reduced band structure notation, but as desired reverses the direction of group velocity $v_g$ to convert spatially diverging to converging rays, see Fig.4A. In particular, the wave with velocity $\vec{v}_g = c_{eff}(\sin(\alpha), \cos(\alpha))$ and wavevector $\vec{k} = k(\sin(\alpha), \cos(\alpha))$, where $k = |\vec{k}| = E/\hbar c_{eff}$, is refracted to $\vec{v}'_g = c_{eff}(\sin(\alpha'), \cos(\alpha'))$ with $\vec{k}' = k'(-\sin(\alpha'), -\cos(\alpha'))$. With $k' = k$ and the wavevector along the lattice (x-axis) being conserved, we obtain $\alpha' = -\alpha$ and recover Snellius law for the negative refraction: $\sin(\alpha)/\sin(\alpha') = n/n' = -1$. We thus expect that π-pulse transfer in the lattice can indeed realize Veselago lensing of the diverging de Broglie waves. To verify that the one-dimensional lattice produces the expected dispersion along a projection of the Dirac cone, we have measured the atomic group velocity along the lattice axis for different values of $k_x$. Corresponding results, see Fig.4b, are in good agreement with the expected dependence.

Fig.4c shows experimental results for a ray tracing simulation of a two-dimensional Veselago lens with cold atoms in the lattice for a diverging pencil of rays with energy $E = 0.104 c_{eff} \hbar \kappa$. To simulate a ray with propagation angle α to the optical axis, the energy gap in the lattice is set to $\Delta E(\alpha) = 2E \cos(\alpha)$ and atoms are prepared with a quasi-momentum $\hbar k_x = (E/c_{eff}) \sin(\alpha)$ in the upper excited Bloch band by a Raman pulse. During the ballistic fall of atoms in the lattice potential, no additional dipole trapping potential is applied in this measurement and one of the lattice beams is swept with a constant chirp rate to compensate for the earth's gravitational acceleration in the atomic rest frame. The data points give the measured centers of the atomic cloud versus the scaled time $\tau = t \cdot c_{eff} \cos(\alpha)/L$, with $L/c_{eff} = 0.25 ms$. The π-pulse transfer, imparted at a scaled time of $\tau = 1$, reverses the atomic paths and focuses the different rays to a single image. This demonstrates negative refraction of the rays, and provides a ray-tracing simulation of Veselago lensing. The variation of the positions where the different rays cross is 2μm, which is order of our measurement accuracy and gives an upper estimate for spherical aberrations of the lens. This value is below the size of the effective de Broglie wavelength $\lambda_{eff} \cong h/\Delta p \cong 7.5 \mu m$ of the velocity selected atomic cloud, indicating that spherical aberrations are small.

*Conclusions*

We have demonstrated negative refraction and Veselago (relativistic) lensing for ultracold atoms in an optical lattice. In the one-dimensional case, we observe refocusing despite the presence of an inhomogeneous external potential. For the case of two-dimensional lensing, a ray-tracing simulation technique was applied. In future, it will be important to study Veselago lensing in optical lattices with higher-dimensional quasirelativistic dispersion[12,14,22-23] to achieve wavepacket refocusing in different spatial dimensions. While in the absence of a bias force the refocusing achieved by the π-pulse transfer occurs at the original position of the wavepacket, forming a temporal version of a Veselago lens, gravity or the Stern-Gerlach force from magnetic gradients can be used to translate the second focus to a spatially separate location. For the future, imaging below the diffraction limit should be examined when imaging objects, such as microscopic mechanical structures, of size smaller than the atomic de Broglie wavelength. Other prospects of π-pulse refocusing in optical lattices with linear dispersion relation include atom interferometric metrology applications, which benefits of the group velocity here being independent of external potentials, as well as its use for novel probes of quantum manybody physics in optical lattices.

**Methods:**

*Experimental Details and Setup*

Our experiment is based on ultracold rubidium atoms in a bichromatic optical lattice potential used to tailor the atomic dispersion relation. Rubidium atoms ($^{87}$Rb) are initially cooled to quantum degeneracy by evaporative cooling in a quasistatic optical dipole trap, formed by a tightly focused horizontally oriented optical beam derived from a $CO_2$-laser operating near 10.6μm wavelength. During the final stage of the evaporation, a magnetic field gradient is activated, which allows us to produce a spin-polarized Bose-Einstein condensate of $6 \cdot 10^4$ atoms in the $m_F=0$ spin projection of the F=1 hyperfine ground state component[24]. A homogeneous magnetic bias of 1.0 G (corresponding to a $\Delta\omega_z \sim 2\pi \times 700$kHz splitting between adjacent Zeeman sublevels) is then applied to remove the degeneracy of magnetic sublevels. By means of Doppler-sensitive Raman transitions, atoms can subsequently be transferred from the $|m_F=0, \hbar k_x\rangle$ to the $|m_F=-1, -2\hbar\kappa + \hbar k_x\rangle$ and $|m_F=-1, 2\hbar\kappa + \hbar k_x\rangle$ states by a 30μs long optical pulse. The transitions are driven by simultaneous irradiation with two copropagating optical beams of frequencies $\omega \pm (\Delta\omega_z + \Delta\omega_d)$ respectively, where $\Delta\omega_d$ corresponds to the Doppler shift of about $2\pi \times 15$kHz of atoms in the corresponding momentum states, and a counterpropagating optical beam of frequency $\omega$. By this atoms are accelerated to "quasi-relativistic" velocities near ±1.1 cm/s in the lattice.

Our method used to create a variable dispersion relation for cold atoms relies on a biharmonic potential with two lattice harmonics of spatial periodicities of λ/2 and λ/4 respectively. Here, λ denotes the wavelength of the driving laser near 783nm wavelength, which is detuned 3nm to the red of rubidium D2-line. Two counterpropagating optical waves derived from this laser are used to generate the fundamental spatial frequency of λ/2, forming a usual standing wave lattice potential. The higher spatial harmonic of periodicity λ/4 is created by virtual four-photon Raman transitions in a three-level scheme with two stable ground states and one electronically excited state. The atoms are irradiated by two copropagating beams of frequencies ω+Δω and ω-Δω and a counterpropagating beam of frequency ω, see Fig. 5. Here, the $m_F$=-1 and the $m_F$=0 spin projections respectively of F=1 are the used ground states of the three-level scheme, while the $5P_{3/2}$ manifold serves as the electronically excited state. We typically use a frequency offset between counterpropagating beams of Δω/(2π) ≅ 1 MHz, which is sufficiently large that two-photon standing wave processes are suppressed. The optical field exchanges momentum with atoms in units of four photon recoils, which is twice that of the corresponding process in a usual standing wave lattice. Correspondingly, the spatial periodicity of the induced lattice potential is a factor two lower than in a standing wave lattice, and reaches λ/4. By combining periodic potentials of λ/2 and λ/4 spatial periodicities, a variable lattice potential can be Fourier-synthesized.

The band structure of the variable lattice along with the energy splitting between the lowest two excited Bloch bands can be calculated as follows, see Refs. 11,18,25 for details. The atomic evolution in the Fourier-synthesized lattice potential is determined by the Hamiltonian

$$H_0 = \frac{\hat{p}}{2m} + V_{\text{lattice}}(x),$$

where $\hat{p}$ denotes the momentum operator and the lattice potential is

$$V_{\text{lattice}}(x) = \frac{V_1}{4}\cos(2\kappa x) + \frac{V_2}{4}\cos(4\kappa x + \varphi).$$

Using a Bloch ansatz to solve the stationary Schrödinger equation, the obtained eigenvalue equation is

$$\sum_{l'} H_{l,l'} \cdot c_{l'}^{(n,k_x)} = E_n^{(k_x)} c_l^{(n,k_x)},$$

where $H_{l,l'}$ can be written in the form

$$H = \begin{pmatrix} \left(\frac{k_x}{\kappa}-2\right)^2 E_r & \frac{V_1}{4} & \frac{V_2}{4}e^{i\varphi} \\ \frac{V_1}{4} & \left(\frac{k_x}{\kappa}\right)^2 E_r & \frac{V_1}{4} \\ \frac{V_2}{4}e^{-i\varphi} & \frac{V_1}{4} & \left(\frac{k_x}{\kappa}+2\right)^2 E_r \end{pmatrix},$$

when we restrict ourselves to the ground and the first two excited Bloch bands. Assuming that $k_x \ll \kappa$, i.e. for atomic quasimomenta $k_x$ near the crossing, and $\Delta E \ll E_r$, an adiabatic elimination of the ground band leads to a reduced 2x2 matrix

$$H_{\text{eff}} = \begin{pmatrix} -\frac{2\hbar^2 \kappa k_x}{m} & \frac{\Delta E}{2} \\ \frac{\Delta E}{2} & +\frac{2\hbar^2 \kappa k_x}{m} \end{pmatrix},$$

in the basis of the eigenstates $|-2\hbar\kappa+\hbar k_x\rangle$ and $|2\hbar\kappa+\hbar k_x\rangle$, where $\Delta E = \left|(V_1^2/32E_r) + (V_2/2)e^{i\varphi}\right|$ denotes the energy splitting between the bands. Using this effective Hamiltonian one can with an additional basis rotation show that the atomic dynamics near the crossing is determined by an effective Dirac equation. As shown in Ref. 18, this also holds in the presence of an additional (spatially slowly varying) external potential. For a finite size of the splitting, atoms will be in a coherent superposition of the eigenstates $|-2\hbar\kappa+\hbar k_x\rangle$ and $|2\hbar\kappa+\hbar k_x\rangle$, meaning that for quasimomenta near the crossing the population in a temporal picture will oscillate between states with positive and negative 'quasirelativistic' velocities. An experimental determination of the group velocity in the upper band versus the atomic quasimomentum has been shown in Fig.4b, which illustrates that the group velocity of atoms near the crossing despite being in the superposition state can well go down to zero values.

The four-photon Raman π-pulse used to transfer atoms between different branches of the dispersion relation in the lattice, between the levels $|m_F=-1,-2\hbar\kappa+\hbar k_x\rangle$ and $|m_F=-1,2\hbar\kappa+\hbar k_x\rangle$ respectively, uses the $|m_F=0,\hbar k_x\rangle$ state as intermediate atomic ground state level. The π-pulse condition can be achieved with a typically 30 µs long optical pulse performed with same optical frequencies as for the Raman beams used for initial state preparation. All required optical frequencies for the variable optical lattice, as well as the Raman beams optical fields used for state preparation and to induce the π-pulse transfer, are derived acoustooptically from a single high power diode laser source. The optical beams after spatial filtering are directed in a vertical geometry (i.e. along the axis of gravity) to the atoms.

*Experimental sequence*

For the one-dimensional Veselago lensing experiment (corresponding to the data shown in Fig.3), atoms are after condensate generation in a tightly focused dipole trap loaded into a dipole potential created with a beam near 10.6µm wavelength and 125µm focal diameter (roughly a factor of 5 larger than the size of the beam used for condensate preparation by evaporative cooling). This beam is irradiated in a counterpropagating direction with respect to the dipole trapping beam used for condensate production, and has an orthogonal polarization. The measured radial trapping frequency of atoms in this weak dipole trap is 70Hz in the presence of the earth's gravitational field. The confining force is so weak that atomic interactions in the dilute atomic cloud are neglible. The velocity width of the expanded condensate atomic cloud is $\cong 0.7\hbar k/m$. The one-dimensional Veselago lensing experiment achieves refocusing of the two spatially diverging atomic rays in the presence of the external potential. When in detail inspecting Fig.3b one finds that the temporal duration required for a refocusing of the wavepackets after the $\pi$-pulse is some 0.2 ms shorter than the period in which the wavepackets expand. This is attributed to a remaining nonlinearity of the bands, which results in slightly different group velocities. In other words, the effective refractive indices in positive and negative energy regions do not exactly match the values $\pm 1$ respectively.

For ray tracing simulation of a two-dimensional Veselago lens (see Fig.4 for corresponding experimental data), after preparation of the rubidium condensate the dipole trapping beam is extinguished and the atoms are left in free fall in the earth's gravitational field. The Raman beams used for preparation and Doppler selection of a particular atomic velocity class along the lattice beams axis are activated 3 ms after release of the atoms from the dipole trapping potential, after which the atomic interaction energy of the dense condensate cloud due to the lowering of density has largely been converted to kinetic energy. The here used low intensity, long (1ms duration) Raman beam pulse selects a narrow slice of $\cong 0.1\hbar k/m$ width from the initial atomic velocity distribution, while the remaining atoms are left in the (undetected) $m_F=0$ spin projection. The lattice beams are subsequently activated. The difference frequency of both vertically upwards and downwards propagating lattice beams and the Raman beams inducing the four-photon π-pulse to reverse the temporal evolution in the course of the Veselago lensing simulation are tuned accousto-optically to yield a spatially stationary lattice field in a frame falling ballistically downwards in the earth's gravitational field. The theory curve in the measurement of the group velocity of Fig.4b (solid line) was derived using the analytic expression

$$v_g(k_x) = \frac{1}{\hbar}\frac{\partial E}{\partial k_x} = c_{eff} k_x / \sqrt{k_x^2 + (\Delta E / 2 c_{eff} \hbar)^2} \ .$$

*Imaging method*

At the end of the experimental sequence, the spatial distribution of atoms in the F=1, $m_F$=-1 ground state level is measured by first transferring this population to F=2, $m_F$=-1 with a 34μs long microwave π-pulse and then pulsing on a laser beam tuned to the F=2 → F'=3 hyperfine component of the rubidium D2-line, recording a shadow image on a CCD camera. From the observed spatial distribution, the width and the center of mass position of the atomic cloud is calculated by fitting a Gaussian distribution to the normalized absorption imaging data.

*Numerical calculations*

We have carried out semiclassical numerical calculations to simulate the atomic center of mass trajectories of the atomic path in the presence of the trapping potential and the lattice, see Figs.3b and 3d. Initially, the dispersion relation in the lattice $E(k_x)$ was calculated by numerically solving the Schrödiger equation for the used experimental parameters. Subsequently, the atomic center of mass trajectory for each path was calculated by solving the classical Euler-Lagrange equations in the presence of the trapping potential and the corresponding atomic dispersion relation in the lattice.

**References:**

1. Veselago, V.G. The electrodynamics of substances with simultaneously negative values of ε and μ. *Sov. Phys. Usp.* **10**, 509-514 (1968).

2. Zhang, X. & Liu Z. Superlenses to overcome the diffraction limit. *Nature Materials* **7**, 435 - 441 (2008).

3. Pendry, J. B., Aubry, A., Smith, D. R. & Maier, S. A. Transformation Optics and Subwavelength Control of Light. *Science* **337**, 549-552 (2012).

4. Pendry, J.B. Negative refraction makes a perfect lens. *Phys. Rev. Lett.* **85**, 3966-3969 (2000).

5. Fang, N., Lee, H., Sun, C. & Zhang, X. Sub–Diffraction-Limited Optical Imaging with a Silver Superlens. *Science* **308**, 534-537 (2005).

6. Xu, T., Agrawal, A., Abashin, M., Chau, K.J. & Lezec, H.J. All-angle negative refraction and active flat lensing of ultraviolet light. *Nature* **497**, 470–474 (2013).

7. Cheianov, V.V., Fal'ko, V.I. & Altshuler, B.L. The Focusing of Electron Flow and a Veselago Lens in Graphene p-n Junctions. *Science* **315**, 1252-1255 (2007).

8. Juzeliūnas, G., Ruseckas, J., Lindberg, M., Santos, L. & Öhberg, P. Quasirelativistic behavior of cold atoms in light fields. *Phys. Rev. A* **77**, 011802 (2008).

9. Eiermann, B. *et al.* Dispersion Management for Atomic Matter Waves. *Phys. Rev. Lett.* **91**, 060402 (2003).

10. Fallani L. *et al.* Optically Induced Lensing Effect on a Bose-Einstein Condensate Expanding in a Moving Lattice. *Phys. Rev. Lett.* **91**, 240405 (2003).

11. Salger, T., Grossert, C., Kling, S. & Weitz, M. Klein Tunneling of a Quasirelativistic Bose-Einstein Condensate in an Optical Lattice. *Phys. Rev. Lett.* **107**, 240401 (2011).

12. Tarruell, L., Greif, D., Uehlinger, T., Jotzu, G. & Esslinger, T. Creating, moving and merging Dirac points with a Fermi gas in a tunable honeycomb lattice. *Nature* **483**, 302-305 (2012).

13. LeBlanc, L.J. *et al.* Direct observation of zitterbewegung in a Bose-Einstein condensate. *New J. Phys.* **15**, 073011 (2013).

14. Soltan-Panahi, P. *et al.* Multi-component quantum gases in spin-dependent hexagonal lattices. *Nature Phys*. **7**, 434-440 (2011).

15. Lamata, L., León, J., Schätz, T. & Solano, E. Dirac Equation and Quantum Relativistic Effects in a Single Trapped Ion. *Phys. Rev. Lett.* **98**, 253005 (2007).

16. Gerritsma, R. *et al.* Quantum simulation of the Dirac equation. *Nature* **463**, 68-71 (2010).

17. Salger, T., Geckeler, C., Kling, S. & Weitz, M. Atomic Landau-Zener Tunneling in Fourier-Synthesized Optical Lattices. *Phys. Rev. Lett.* **99**, 190405 (2007).

18. Witthaut, D., Salger, T., Kling, S., Grossert, C. & Weitz, M. Effective Dirac dynamics of ultracold atoms in bichromatic optical lattices. *Phys. Rev. A* **84**, 033601 (2011).

19. Hahn, E. L. Spin Echoes. *Phys. Rev.* **80**, 580-594 (1950).

20. Berman, P.R. *Atom Interferometry* (Academic Press, San Diego, 1997).

21. Cronin, A. D., Schmiedmayer, J. & Pritchard, D. E. Optics and interferometry with atoms and molecules. *Rev. Mod. Phys.* **81**, 1051-1129 (2009).

22. Zhu, S.-L., Wang, R. & Duan, L.-M. Simulation and Detection of Dirac Fermions with Cold Atoms in an Optical Lattice. *Phys. Rev. Lett.* **98**, 260402 (2007).

23. Lee, K.S. *et al.* Ultracold fermions in a graphene-type optical lattice. *Phys. Rev. A* **80**, 043411 (2009).

24. Cennini, G., Ritt, G., Geckeler, C. & Weitz, M. All-Optical Realization of an Atom Laser. *Phys. Rev. Lett.* **91**, 240408 (2003).

25. Ritt, G., Geckeler, C., Salger, T., Cennini, G. & Weitz, M. Fourier Synthesis of Conservative Atom Potentials. *Phys. Rev. A* **74**, 063622 (2006).

**Acknowledgements**

We acknowledge discussions with L. Santos and J. Klaers. Financial support from the DFG (We1748-20) is acknowledged.

**Competing financial interests**

The authors declare that they have no competing financial interests.

**Author contributions**

M. L. and C. G. conducted the experiment and analysed the data. All authors contributed to the writing of the paper and to the interpretation of the results. M. W. planned the project.

**Figure captions:**

**Figure 1 | Schematic of a Veselago lens in the *x-z* plane.** A diverging pencil of rays is focused upon entry into a medium with refractive index n=-1 by negative refraction. The negative refraction can be described by Snell's law when noting that the refractive index at the surface changes sign from positive to negative.

**Figure 2 | Atomic dispersion in lattice.** The figure shows the dispersion relation of rubidium atoms in the bichromatic optical lattice for $V_1$=2.1$E_r$, $V_2$=0.28$E_r$ and a relative phase between lattice harmonics of φ=180° (solid) and φ=0° (dashed). In the former case, the splitting between the first (red solid line) and the second (blue solid line) excited Bloch band vanishes and the dispersion near the crossing becomes linear, i.e. relativistic. The zero of the energy scale is here set to the position of the crossing. Transitions between the branch of the dispersion relation with E>0, corresponding to the case with collinear wavevector and group velocity, and the regime with E<0, for which wavevector and group velocity are antiparallel, can be induced by four-photon Raman π-pulses.

**Figure 3 | Veselago lensing in one dimension. a**, Schematic of the relevant part of the atomic dispersion for a vanishing splitting between first (red) and second (blue) excited Bloch bands with the corresponding group velocities indicated. **b,** Temporal sequence of

absorption images for the case of such an ultrarelativistic dispersion in the lattice. After 1.3ms a π-pulse is applied, leading to a refocusing. The false-colour images give the optical column density of atoms in the $m_F$=-1 spin projection, as detected by a combination of a short microwave pulse transferring the population to the upper (F=2) hyperfine ground state and then applying a standard absorption imaging technique. The used temporal bin widths in the left panel is 100μs, and the white lines are calculated trajectories for $k_x$={±0.05,±0.1,±0.2}$\kappa$ respectively. **c,** Variation of the atomic cloud width versus time. **d,e,** Corresponding data for a nonvanishing splitting between the Bloch bands, which leads to less perfect refocusing. For the data shown in the left panel, ΔE=0.28 $E_r$. The shown error bars are given by the standard deviation of the mean, as determined by Gaussian fits to the atomic density distribution for three corresponding data sets.

**Figure 4 | Veselago lensing in the *x-z* plane. a,** Representation of negative refraction, with corresponding parallel and antiparallel alignments of wavevector and group velocity respectively before and after the refraction. **b,** Measured group velocity $v_x$ along the direction of the lattice beams for $\Delta E = 0.208 E_r$ versus atomic quasi-momentum $\hbar k_x$. The data points are overlayed with the expected dependence $v_g(k_x) = \frac{1}{\hbar}\frac{\partial E}{\partial k_x}$, shown by the solid line. **c,** Ray tracing simulation of two-dimensional Veselago lensing. The data points give the measured position of the atomic cloud, as derived by absorption imaging, versus the scaled time $\tau = t \cdot \cos(\alpha)/0.25 ms$ for different angles α of the simulated trajectory. At $\tau = 1$, a π-pulse is applied, leading to a redirection of paths to a second focus. The shown error bars are given by the standard

deviation of the mean, as determined by Gaussian fits to the atomic density distribution for three corresponding data sets.

**Figure 5 | Subwavelength optical lattice.** Virtual four-photon Raman process leading to an optical lattice potential with spatial periodicity $\lambda/4$, as the higher harmonic of the used biharmonic potential for cold atoms.

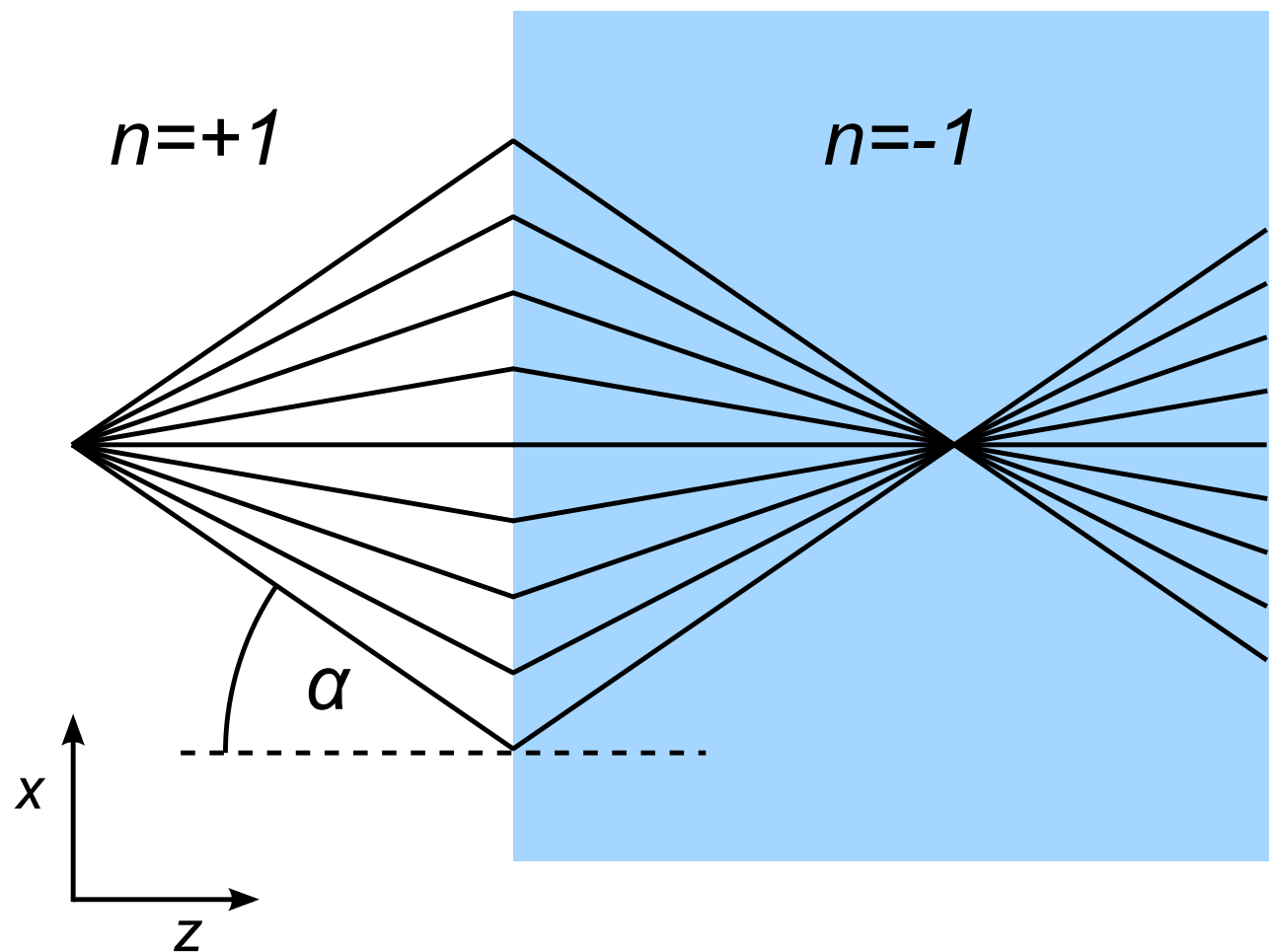

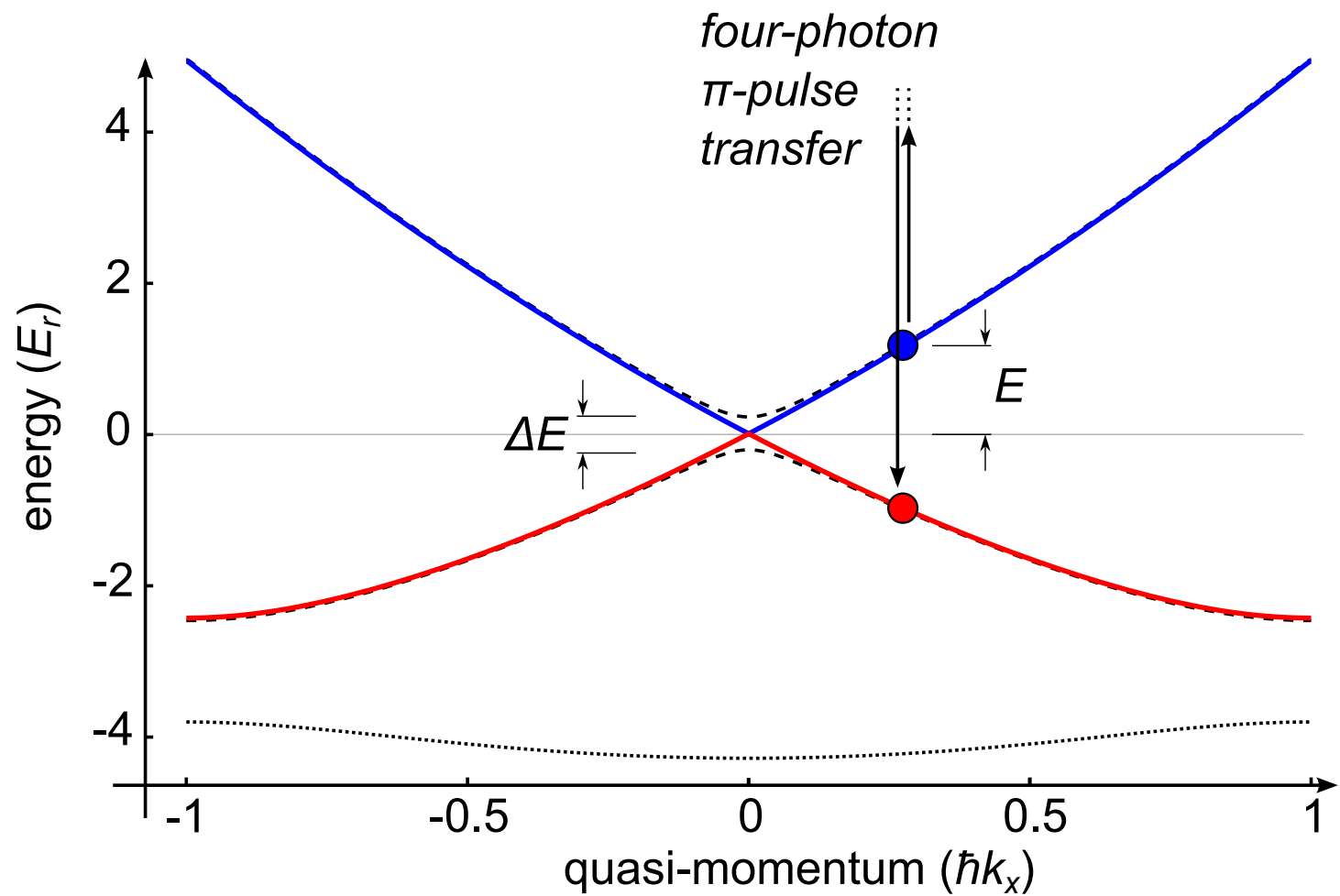

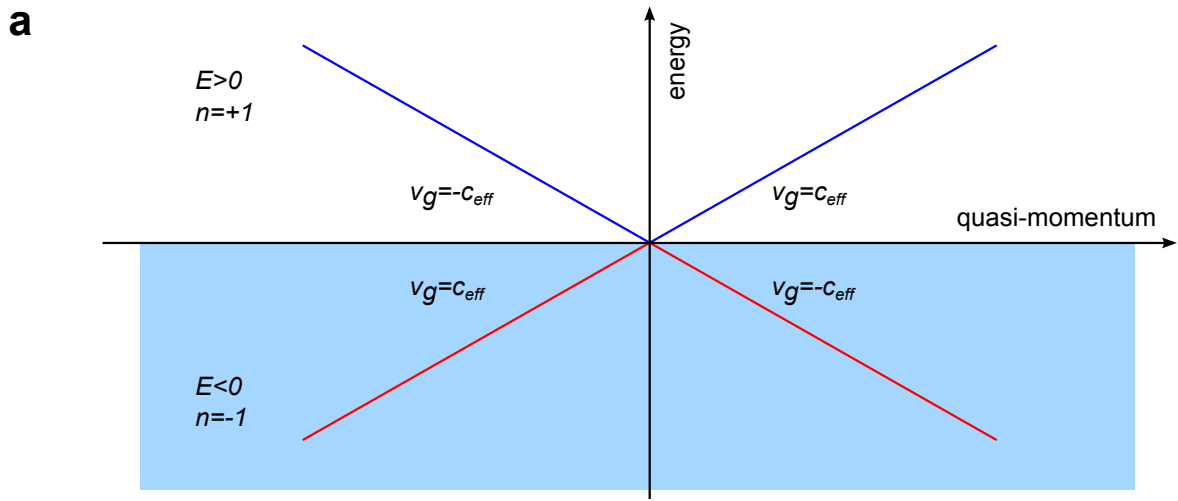
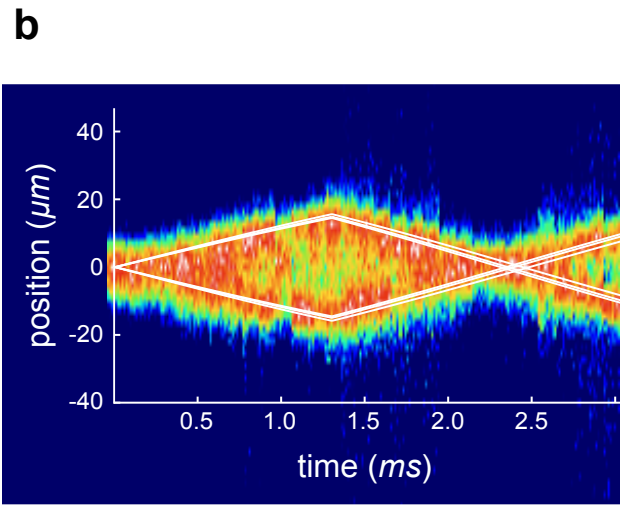
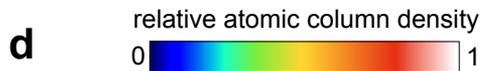
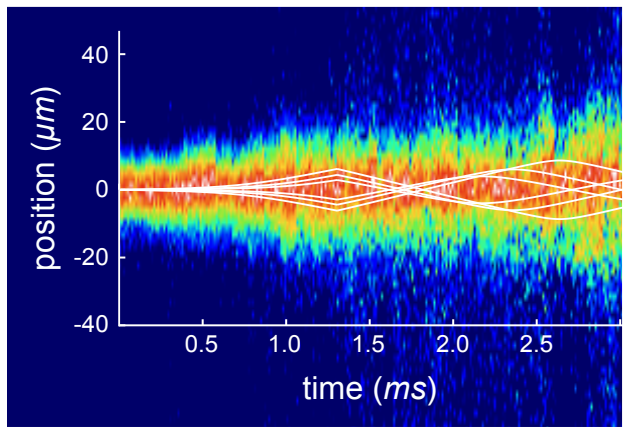
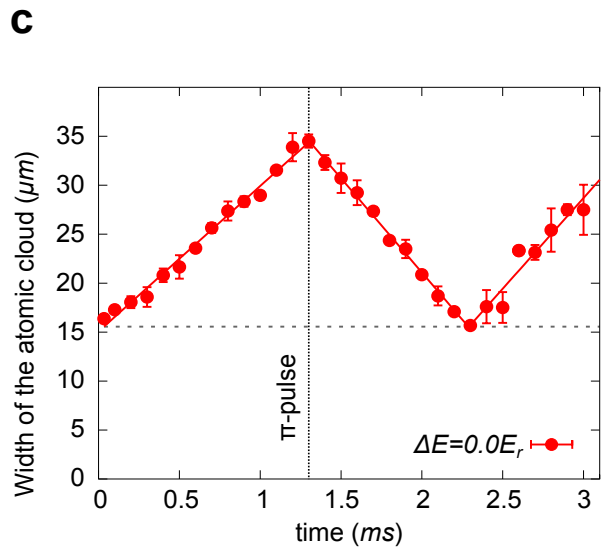
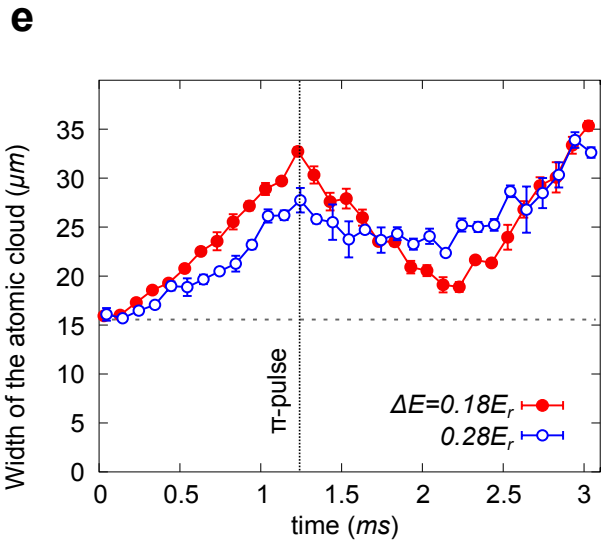

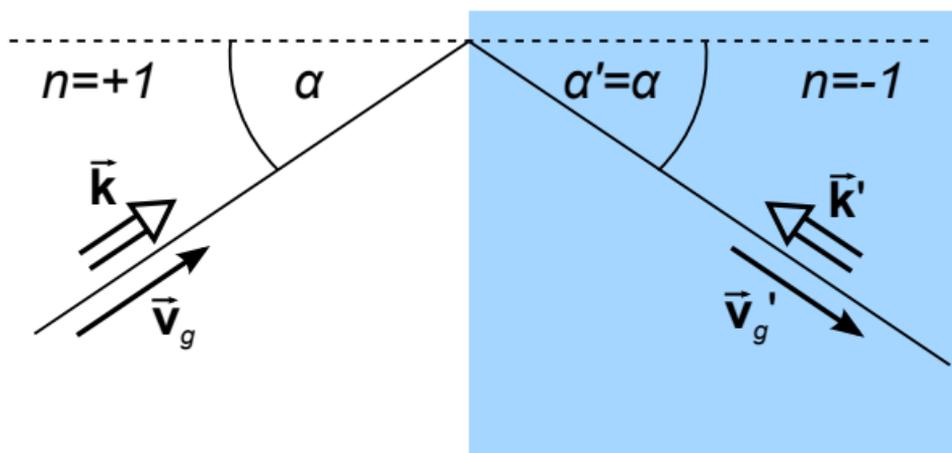
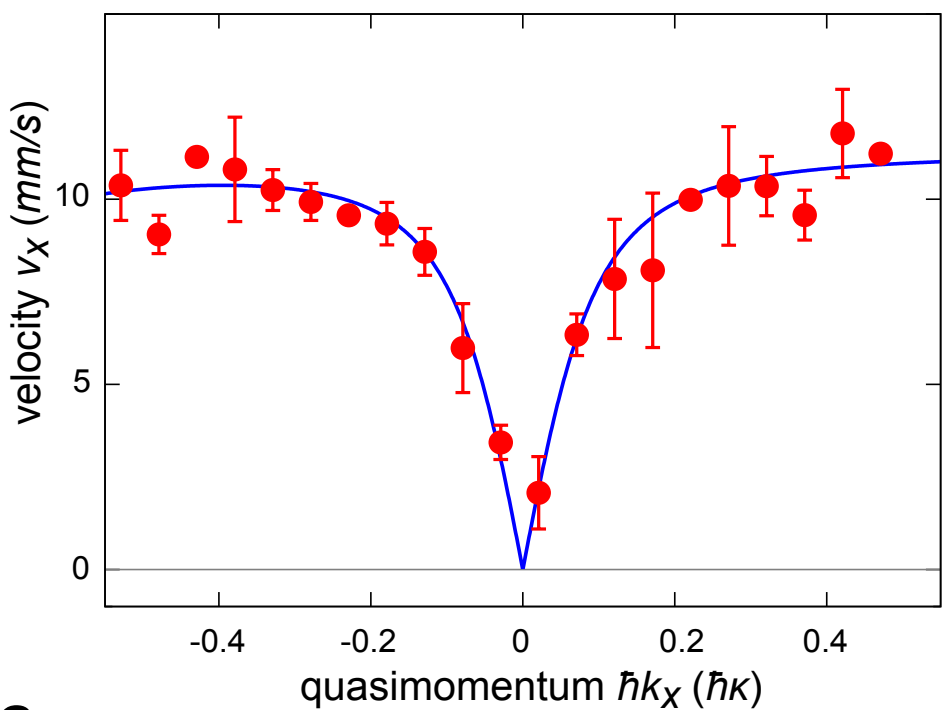
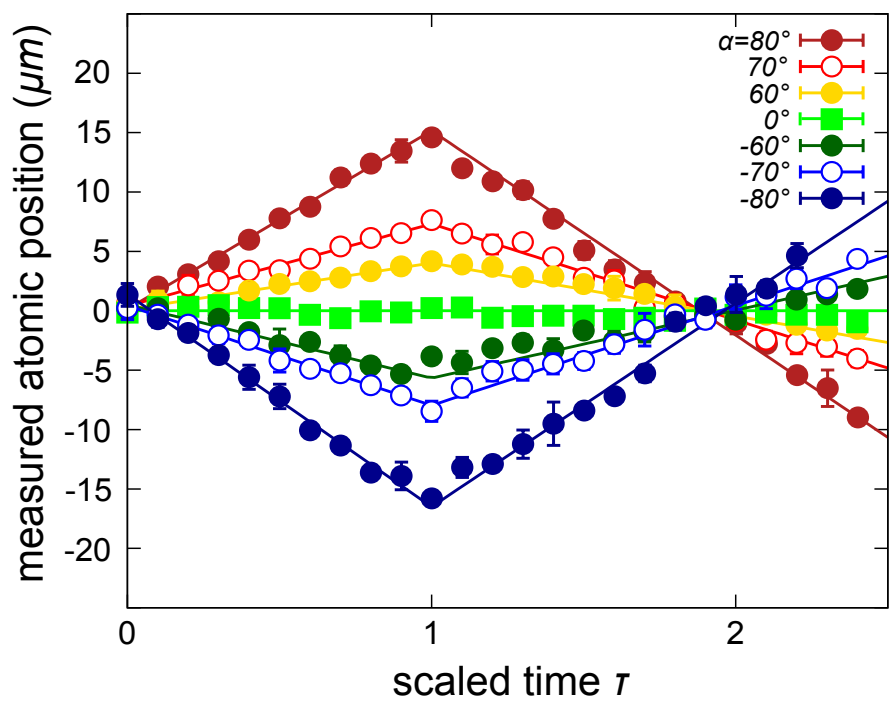

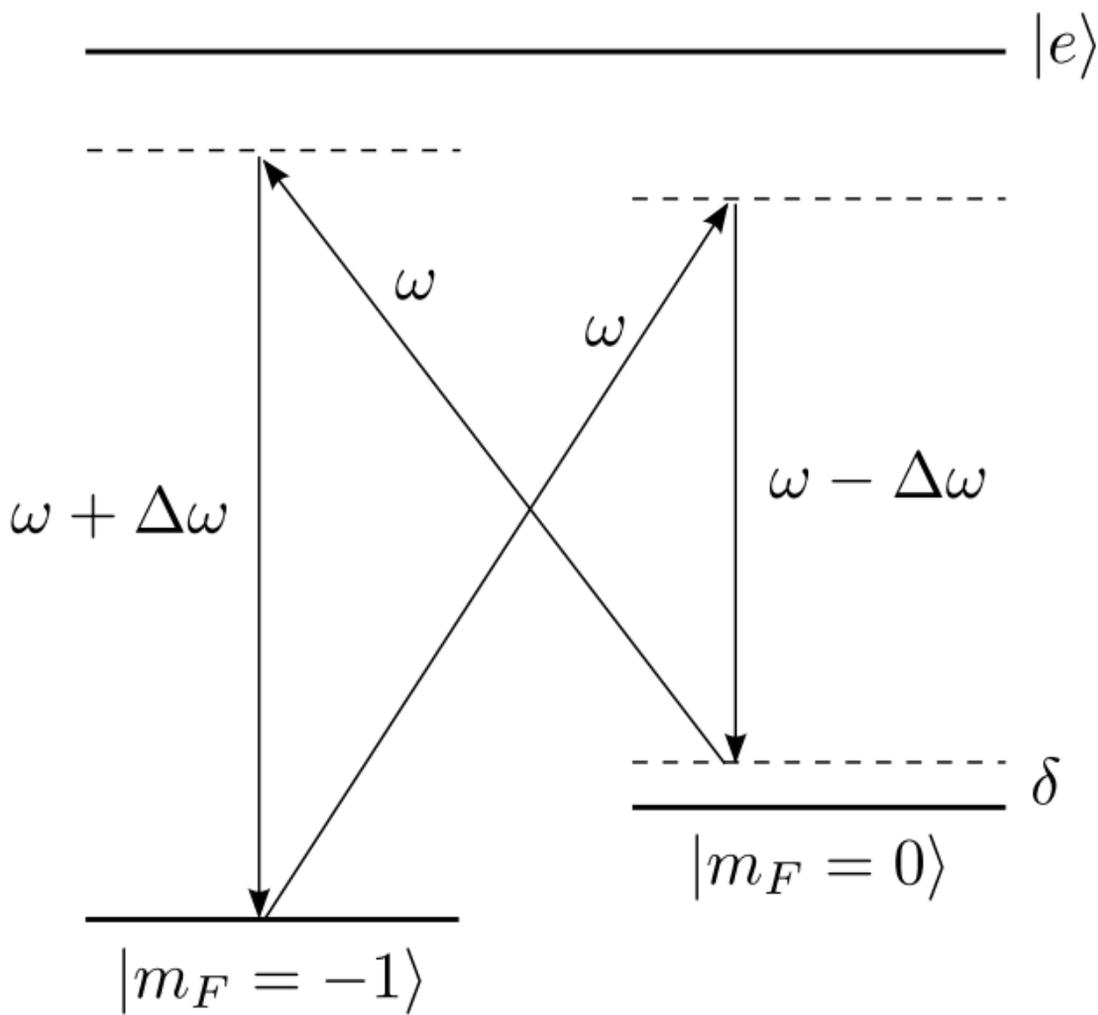